\pdfoutput=1
\documentclass[aip,apl,reprint]{revtex4-1}
\usepackage{graphicx}
\usepackage{dcolumn}
\usepackage{bm}
\usepackage{epstopdf}
\usepackage{color}
\draft 
\begin{document}
\title{Current-injection quantum-entangled-pair emitter using droplet epitaxial quantum dots on GaAs(111)A} 
%
%
%
\author{Neul~Ha}
%
\author{Takaaki Mano}
\author{Takashi Kuroda}
\email[Author to whom correspondence should be addressed: ]{kuroda.takashi@nims.go.jp}
\author{Yoshiki Sakuma}
\author{Kazuaki Sakoda}
\affiliation{National Institute for Materials Science, 1-1 Namiki, Tsukuba 305-0044, Japan}


\date{\today}

\begin{abstract}
A source of single photons and quantum entangled photon pairs is a key element in quantum information networks. Here, we demonstrate the electrically driven generation of quantum entangled pairs using a naturally symmetric GaAs quantum dot grown by droplet epitaxy. Coincidence histograms obtained at a temperature of 10~K reveal the generation of quantum entangled pairs that have a fidelity to the Bell pairs of $0.71 \pm 0.015$, much beyond the classical limit. The quantum nature of the emitted pairs is conserved at temperatures of up to 65~K, and is essentially limited by the charge carrier confinement in the present dot system. Our study offers a guideline for the fabrication of quantum entangled pair sources suitable for practical use. 
\end{abstract}

%
\maketitle 
\noindent\textit{\textbf{Introduction. }}%
Quantum entanglement is an essential resource for the implementation of quantum information processing. In a quantum network, for example, various quantum devices are expected to be linked via a photonic stream that serves as flying qubits. Entangled photon pairs transmitted in the network create nonlocal correlations between quantum devices, and enable complex processing that is not available with a classical architecture. 
A photon emitting device is therefore a key element of a quantum network. %

The use of a semiconductor quantum dot (QD) as an entangled pair source was initially proposed in 2000 \cite{Benson_PRL00}. Despite the concept being straightforward and analogous to that of an atomic cascade \cite{Aspect_PRL81}, experimental implementation has been problematic \cite{Santori_PRB02} due to the presence of optical anisotropy inherent in commonly used dots %
\cite{Gammon_science96, Bayer_PRL99, Seguin_PRL05}. 
Several postproduction techniques have already been developed to recover the QD isotropy %
\cite{Stevenson_Nature06, Akopain_PRL06, %
Hafenbrak_NewJP_07,*Keil_NatComm17,%
Muller_PRL09, Dousse_Nature10, Ghali_Ncom12,%
Trotta_PRL12, *Huber:2018jz}. 
An alternative way to achieve an isotropic dot is to use a $C_{3v}$ symmetric \{111\} crystallographic surface as a growth substrate. 
The triangular arrangement of atoms on \{111\} surface eliminates a source of structural and wavefunction elongation, and leads to near-perfect isotropy with a vanishing bright exciton splitting \cite{Singh_PRL09, Schliwa_PRB09}. 
Although conventional Stranski-Krastanov QD growth is prohibited on a \{111\} surface, other techniques, such as pyramidal selective etching %
\cite{Juska_Nphoto13, Chung_Nphoto16}, 
core-shell nanowire growth %
\cite{Versteegh_Ncom14, Jons_SciRep2017}, 
and droplet epitaxy %
\cite{Mano_APEx10,Kuroda_PRB13, %
Ha_APL14,*Liu_PRB14,*Ha_APEx16, %
BassoBasset_Nano17}, 
have been used to create isotropic dots on \{111\} for the generation of quantum entangled pairs. Among these techniques, droplet epitaxy is a self-assembled QD growth technique, and is thus suitable for large-scale device integration. %

In this work, we demonstrate the electrically driven operation of entangled pair emission from droplet epitaxy GaAs QDs. Previous studies on the droplet QD source have focused on optical excitation with the aid of an external laser. This scheme is easy to implement, but not favorable for practical applications. This drawback can be overcome by constructing an electrically driven source of entangled pairs 
\cite{Salter:2010gj,*Varnava_npj16,%
Zhang_NatComm2015}. Here, we develop a photon emitting device comprising a naturally symmetric QD grown on a Ga-rich GaAs(111)A substrate, encapsulated in a \textit{p-i-n} diode structure. Thanks to our careful optimization of the growth of doped barrier layers along the [111] direction, our entangled-pair emitters can work at temperatures of up to 65~K, and are simply limited by the charge carrier confinement of the GaAs/AlGaAs hetero-system.  %

\begin{figure}
\includegraphics[width=7.5cm]{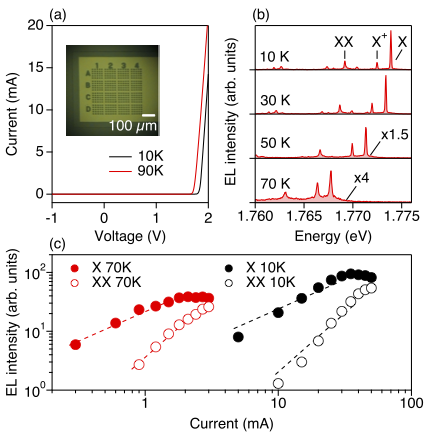}
\caption{\label{fig_spctr}(a) I-V dependence measured at 10 and 90~K (sample~A). The inset shows the top view of our device. (b) EL spectra of a single GaAs QD measured at 10, 30, 50, and 70~K (sample~A). (c) X and XX intensities as a function of bias current at 10 and 70~K (sample~A). Broken lines behind the data points are linear and quadratic dependences drawn to guide the eye. }
\end{figure}

\textit{\textbf{Sample preparation. }}%
The QD diode sample is grown on \textit{n}$^+$ doped GaAs(111)A using a solid source molecular beam epitaxy machine. After depositing Si-doped \textit{n}-type GaAs buffer ($n=1\times10^{18}$~cm$^{-3}$), we grow a Si-doped Al$_{0.25}$Ga$_{0.75}$As layer ($n=5\times10^{17}$~cm$^{-3}$, 200~nm thick), which serves as an electron supplying layer. Here, we optimize the aluminum composition of the AlGaAs barrier at 0.25, which makes \textit{n}-type conduction possible even at low temperatures, while reasonably high barrier height is maintained \cite{Ishibashi_JJAP82}. 
The QD layer is separated from the \textit{n}-doped (\textit{p}-doped) region by a 120~nm (80~nm) thick intrinsic Al$_{x}$Ga$_{1-x}$As layer ($x=0.25$). 
In addition, we insert an atomically thin Al$_{x}$Ga$_{1-x}$As layer ($x=0.35$, 1~nm~thick) under the QD layer to avoid the occurrence of the local droplet etching effect during the droplet formation process \cite{Wang_APL07, Heyn_PRB11}. 

For the QD formation, 0.05 of a monolayer (ML) of gallium is supplied at a growth rate of 0.01 ML/s at 450~$^{\circ}$C, yielding the formation of gallium droplets on AlGaAs(111)A with a density of $4 \times 10^8$~cm$^{-2}$. Then, an As$_4$ flux ($5\times10^{-6}$~Torr) is supplied at 200~$^{\circ}$C for the crystallization of GaAs dots from gallium droplets, followed by \textit{in vacuo} annealing at 500~$^{\circ}$C. With these conditions, GaAs QDs have a disk-like shape with an average base diameter of 40~nm and a height of $1.0$~nm (see supplementary material for an atomic microscope image of the QD surface). 

After capping the QDs with un undoped AlGaAs barrier, C-doped \textit{p}-type Al$_{0.25}$Ga$_{0.75}$As ($p=5\times10^{17}$~cm$^{-3}$, 200~nm thick), highly C-doped Al$_{0.25}$Ga$_{0.75}$As ($p=2\times10^{18}$~cm$^{-3}$, 20~nm thick), and C-doped GaAs ($p=2\times10^{19}$~cm$^{-3}$, 20~nm thick) layers are formed at 520~$^{\circ}$C. A top ohmic electrode with 10~$\mu$m square open windows is fabricated with photolithography and Ti/Pt/Au deposition. Finally, chemical etching is performed to form a mesa structure in order to restrict the carrier flow region. A top view of this device, denoted sample~A, is shown in the inset of Fig.~\ref{fig_spctr}(a). We also study a similar diode structure with a slightly different layer sequence, denoted sample~B, and its growth is described in the supplementary material. 

\textit{\textbf{Optical setup. }}%
Forward DC bias is applied between the top and bottom electrodes, so that charge carriers are continuously injected into the GaAs QDs. Electroluminescence (EL) signals are collected by using a micro-objective lens with a numerical aperture of 0.8, and then passed through a wave retarder and a Glan prism, which serves as a polarization analyzer. Then, the photon beam is fed into a monochromator equipped with a silicon avalanche photodiode. We count three photon channels simultaneously, i.e., biexciton (XX) and exciton (X) photons with a given polarization state, and X photons with an orthogonal polarization state. This setup enables us to suppress the effect of detection-yield fluctuation during measurement. The number of coincidences is analyzed using a fast-response time-to-digital converter.

\textit{\textbf{Results and discussion. }}%
Figure~\ref{fig_spctr}(a) shows the current-voltage (I-V) characteristics of our device (sample A). It reveals an asymmetric response with a threshold at a forward bias, which is a characteristic of a \textit{p-i-n} diode structure. The threshold voltage is observed to be lower at 90~K than at 10~K due to the thermal activation of free carriers. 

Figure~\ref{fig_spctr}(b) shows the EL spectrum of a single GaAs QD. Three emission lines are observed, and assigned as charge neutral X, positively charged X (X$^+$), and charge neutral XX, from the higher energy side. The EL dependence on temperature is investigated, where the bias current is adjusted so that the X intensity reaches $\sim80$~\% of its saturation value. With increasing temperature, all the spectral lines move to the lower energy side along with the temperature shift in the band gap. They are broadened simultaneously, and accompanied by the acoustic phonon band at temperatures as high as 70~K\cite{Abbarchi_JAP08}. Note that significant X and XX lines are observed even at 70~K (although the intensity is $\sim 1/4$ of that at 10~K). Stable QD emission at such high temperatures is not common in standard photo-injection devices using GaAs dots, implying that effective carrier injection is achieved in the \textit{p-i-n} diode. 

Figure~\ref{fig_spctr}(c) shows the dependences of the X and XX intensities on bias current. Linear and quadratic dependences are confirmed in the low current regions for X and XX signals, respectively, supporting their spectral identification. 
We find that the bias current needed to reach the saturation intensity is more than ten times higher at 10~K than at 70~K. This is due to the quenching of the active carrier concentration at low temperatures, as discussed in the supplementary material. 

\begin{figure}
\includegraphics[width=7.5cm]{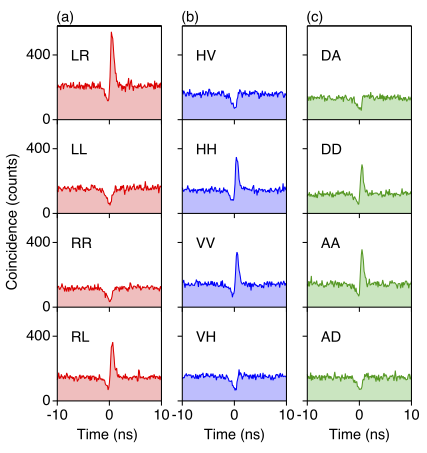}
\caption{\label{cncdnc_set}Coincidence histograms between XX and X photons emitted from a GaAs QD for different polarizations at 10~K (sample B). 
The two-photon projection settings (such as LR) are indicated by the first letter for XX photons and the second letter for X photons. The bias current is set so that the EL intensity almost reaches its saturation value ($I=40$~mA). The histograms are integrated with a time bin of 128~ps.}
\end{figure}

The results of photon correlation measurements at 10~K are summarized in Fig.~\ref{cncdnc_set}, where L, R, H, and V are the left-hand circular, right-hand circular, linear horizontal, and vertical polarizations, respectively; D is linear diagonal with a polarization axis tilted by 45~$^\circ$ from H, and A is antidiagonal where A$\perp$D. 
The top panel in Fig.~\ref{cncdnc_set}(a) is a coincidence histogram for L-polarized XX photons and R-polarized X photons (denoted LR). The central coincidence peak has an asymmetric shape that consists of a dip (negative peak) at $t \leq 0$ and a positive peak at $t \geq 0$. Note that coincidence signals at positive times are counted for the detection of an XX photon followed by that of an X photon. Thus, the measured asymmetric evolution in the coincidence traces confirms the XX-X radiative cascade. The XX and X photons that have a polarization combination of LR are clearly correlated, resulting in a higher probability than that for detecting uncorrelated photons. The peak disappears for LL and RR, while it recovers for RL. 

Figures~\ref{cncdnc_set}(b) and \ref{cncdnc_set}(c) show coincidence histograms for linear polarizations. A positive correlation appears for parallel polarization (\textit{e.g.} HH and DD), while it disappears for perpendicular polarization (\textit{e.g.} HV and DA). These polarization correlations, which are independent of the choice of projection basis, suggest that the measured two-photon wavefunction is approximated by one of the maximally entangled (Bell) states, $\arrowvert \Psi^+ \big \rangle = \{ \vert \mathrm{HH} \rangle + \vert \mathrm{VV} \rangle \} /\sqrt{2} \equiv \{ \vert \mathrm{RL} \rangle + \vert \mathrm{LR} \rangle \} /\sqrt{2}$. %
It should be noted that the coincidence dip at $t \leq 0$ does not reach zero, although we expect no probability of observing an X photon followed by an XX photon (reversed cascade). This non-ideal signature is due to the finite resolution of the coincidence setup. Our photon detector has a timing jitter of $\sim 400$~ps close to the emission lifetime of GaAs dots ($\sim600$~ps). Hence, the sharp antibunching dip is smoothed out, and observed to stay at a finite value at $t=0$. This smoothing effect also influences the precise evaluation of positive correlation intensities, and makes it difficult to determine the degree of quantum entanglement. 

\begin{figure}
\includegraphics[width=7.5cm]{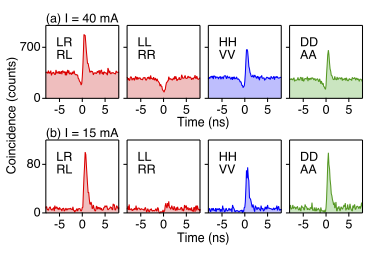}
\caption{\label{cncdnc_LowPower} Comparison of coincidence histograms at 10~K measured with bias currents of (a) 40~mA and (b) 15~mA (sample B). In each panel we plot the sum of the coincidence traces for two equivalent projection settings, such as LR and RL. 
}
\end{figure}

To quantify the correlation degree, we measure coincidence signals with a reduced current injection, where we expect to suppress the probability of counting accidental (uncorrelated) photons \cite{Oohata_PRL07,kuroda_PRB09}. Figure~\ref{cncdnc_LowPower} reveals the impact of current injection \textit{I} on the coincidence trace. When $I=40$~mA, close to a saturation current, we observe the correlation peaks superimposed on the high backgrounds associated with accidental coincidence (Fig.~\ref{cncdnc_LowPower}(a)). When $I=15$~mA, the background is greatly reduced, although a significant coincidence peak remains (Fig.~\ref{cncdnc_LowPower}(b)). Thanks to these nearly background-free traces, we can evaluate the polarization degree for each projection setting, i.e., $C_{\mathrm{RL}} = 0.70 \pm 0.03$, and $C_{\mathrm{HV}} \approx C_{\mathrm{DA}} = 0.56 \pm 0.03$, where the polarization degree is defined as $C=\left\vert (n_{\|}-n_{\perp})/(n_{\|}+n_{\perp}) \right\vert$, and $n_\|$ ($n_\perp$) is the coincidence number 
for a co-polarized (cross-polarized) setting. Note that we observe a lower correlation in linear polarization than circular polarization. This is a typical signature of dephasing induced by nuclear spin noise. Using the above set of $C$ values we determine the fidelity to the Bell state as $f = 0.71 \pm 0.015$, where we adopt the simple expression $f=(1+C_{\mathrm{RL}}+C_{\mathrm{HV}}+ C_{\mathrm{DA}})/4$ \cite{Hudson_PRL07}. The measured fidelity to Bell pairs is higher than the classical upper limit of 0.5 by more than ten times the standard deviation. Thus, we confirm that our diode generates quantum entangled pairs. 

\begin{figure}
\includegraphics[width=7.5cm]{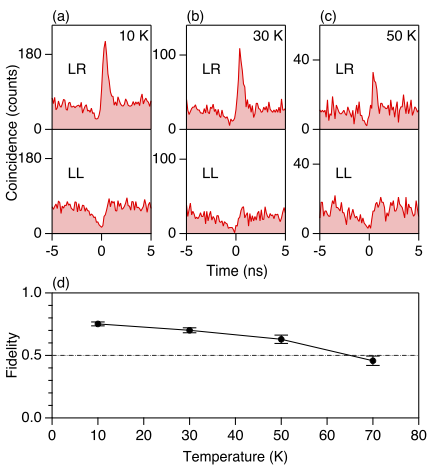}
\caption{\label{cncdnc_highTemp}Temperature dependence of coincidence histograms for circular polarization combinations measured at (a) 10~K, (b) 30~K, and (c) 50~K (sample A). (d) The fidelity to the Bell state of emitted photons as a function of operation temperature.}
\end{figure}

We also study the impact of increasing temperature on device performance. Figures~\ref{cncdnc_highTemp}(a), \ref{cncdnc_highTemp}(b), and \ref{cncdnc_highTemp}(c) show coincidence histograms for co- and cross-circular polarizations measured at different temperatures. In this measurement we adjust the bias current so that the X intensity is kept at $\sim80$~\% of its saturation value. With increasing temperature, the cross-circular (LR) peak height decreases, but the peak signature is apparent even at 50~K. A similar correlation is measured for the other polarization basis, and shown in the supplementary material. 
Figure~\ref{cncdnc_highTemp}(d) shows the fidelity to the Bell state as a function of temperature, where we integrate the coincidence counts over a time interval of 512~ps, which is similar to the emission lifetime. The fidelity value decreases with temperature, and crosses the classical limit of 0.5 at $\sim 65$~K, which serves as a guideline for the maximum operation temperature of our device. 

We attribute the correlation reduction with temperature to the shallow confinement of the charge carriers in our GaAs/AlGaAs QD system. Note that the confinement depth of charge carriers is estimated to be around 50~meV in the present dot. This weak confinement means that charge carriers at high temperatures easily escape from dots. As the time scale of carrier escape/injection becomes comparable to that of the XX-X emission cascade, the probability of observing uncorrelated photons increases, and the correlation degree decreases. The robustness against increasing temperature could therefore be improved by using a heterostructure material that has a higher quantum confinement, as recently demonstrated in a telecommunication-wavelength QD system \cite{Muller_Naturecom18}. 

\textit{\textbf{Conclusion. }}%
We successfully fabricated an electrically driven entangled pair source using droplet epitaxy GaAs dots embedded in a \textit{p-i-n} diode. Thanks to the high structural symmetry of dots grown on (111)A, we confirmed a clear polarization correlation whose degree is well beyond the classical limit even with a D.C. current injection. The quantum nature of the emitted photon pairs is conserved at operation temperatures up to 65~K, which is essentially limited by the shallow confinement of charge carriers in GaAs/AlGaAs hetero-systems. Thus, the application of a large offset hetero-system to the QD base material is a potential route to achieving higher operating temperatures. 

See supplementary material for an atomic microscope image of the QD surface, the detailed layer sequence of the studied diodes, and a complete set of coincidence histograms for all polarization combinations at different temperatures. 

We acknowledge the support of a Grant-in-Aid from the Japan Society for the Promotion of Science. 


\bibliography{entangleLED.bib}

\end{document}